\documentclass{aastex}
\usepackage{emulateapj5}
\usepackage{apjfonts}
\usepackage{epsf}

\journalinfo{THE ASTROPHYSICAL JOURNAL}
\slugcomment{Accepted by the Astrophysical Journal on 12/July, 2002.}
\shorttitle{Accretion Rates and SEDs of BL Lac Objects}
\shortauthors{Wang, Staubert, \& Ho}

\def\MR{M_{\rm R}}

\def\llpk{\log L_{\rm pk}}
\def\lll{\log L_{\rm lines}}
\def\lline{L_{\rm lines}}
\def\lnupk{\log \nu_{\rm pk}}
\def\mg{Mg\,{\sc ii}}
\def\mbh{m_{\rm BH}^{}}

\def\oo{[O\,{\sc ii}]}
\def\ooo{[O\,{\sc iii}]}
\def\pknu{\nu_{\rm pk}}
\def\pkl{L_{\rm pk}}

\newdimen\digitwidth      
\setbox1=\hbox{0}         
\digitwidth=\wd1          
\catcode`"=\active        
\def"{\kern\digitwidth}

\begin{document}

\title{THE ACCRETION RATES AND SPECTRAL ENERGY DISTRIBUTIONS OF BL LACERTAE
OBJECTS}

\author{Jian-Min Wang\altaffilmark{1,2,3},
R\"udiger Staubert\altaffilmark{1}
and Luis C. Ho\altaffilmark{4}}

\altaffiltext{1}{Institut f\"ur Astronomie und Astrophysik, Abt. Astronomie,
Universit\"at T\"ubingen, Sand 1, D-72076 T\"ubingen, Germany;
wang@astro.uni-tuebingen.de.}

\altaffiltext{2}{Laboratory for High-Energy Astrophysics, Institute of
High-Energy Physics, CAS, Beijing 100039, P.~R. China.}

\altaffiltext{3}{Alexander von Humboldt Fellow.}

\altaffiltext{4}{The Observatories of the Carnegie Institution of Washington,
813 Santa Barbara Street, Pasadena, CA 91101-1292.}

\begin{abstract}
We investigate the relationship between accretion rates and the spectral energy
distributions (SEDs) of BL Lac objects, using a sample of objects for which
published information on the host galaxies, emission-line luminosities, and 
peak frequencies and luminosities of their SEDs are available. The sample is 
composed of 43 BL Lac objects which have a relatively continuous distribution 
of peak frequencies. Under the assumption that the observed emission lines 
are photoionized by the central accretion disk, we use the line luminosities to 
estimate the accretion luminosities and hence accretion rates.  We 
find that low frequency-peaked BL Lac objects (LBLs) span a wide range of
accretion rates, whereas high frequency-peaked BL Lac objects (HBLs) cover a
more restricted range of lower values.  There appears to be a continuous
distribution of accretion rates between the two subclasses of BL Lac objects.
We find that the peak frequency of the SED, $\pknu$, correlates with the
accretion rate, approximately with the form $\pknu\propto \Lambda^{-3}$ in
HBLs and $\pknu \propto \Lambda^{-0.25}$ in LBLs, where
$\Lambda \equiv \lline/c^2$. The peak luminosity of the SED is also
correlated with $\Lambda$.  These results suggest that the accretion
rate influences the shape of the SED in BL Lac objects.  They also support
models which couple the jet and the accretion disk.  We present a physical
scenario to account for the empirical trends.
\end{abstract}


\section{INTRODUCTION}

The spectral energy distributions (SEDs) of BL Lac objects can be largely
characterized by two broad peaks. The low and high energy peaks are commonly
attributed to synchrotron emission and inverse Compton scattering,
respectively.  Depending on the relative strengths of the two peaks, it is
now customary to divide BL Lac objects into low frequency-peaked (LBL) and
high frequency-peaked (HBL) sources (Giommi \& Padovani 1994). Typically, HBLs
are less variable, less polarized, and contain less dominant radio cores than
LBLs (e.g., Laurent-Muehleisen et al. 1993; Perlman \& Stocke 1993; Jannuzi,
Smith \& Elston 1994). However, the physical origin of the differences between
the two classes of BL Lac objects remains a matter of debate.

According to the unification scheme for BL Lac objects, LBLs and HBLs are
observed at different viewing angles (see review by Urry \& Padovani 1995).
On the other hand, Sambruna, Maraschi, \& Urry (1996) have shown that the
typical multiwaveband SED of an HBL cannot be produced from the spectrum of an 
LBL simply by changing the viewing angle alone. They suggested that other
physical parameters, such as the magnetic field, may be different in the two
classes.  New X-ray and radio surveys, such as those by Perlman et al. (1998),
Laurent-Muehleisen et al. (1998), and Caccianiga et al. (1999), have shown
that the distribution of peak frequencies in the SEDs of BL Lac objects is
more uniform than previously thought.  What is the main physical parameter (or
set of parameters) that control the SEDs of BL Lac objects?

Recent observations of the host galaxies of BL Lac objects provide new clues to
understanding the physical nature of these systems.  Urry et al. (2000) and
Scarpa et al. (2000) systematically studied the morphologies of BL Lac objects
using the {\it Hubble Space Telescope (HST)}\ and found that there is no
significant difference between the host galaxies of HBLs and LBLs. Most of the
hosts are normal, giant elliptical galaxies, with no obvious evidence for
ongoing mergers or strong interactions with other galaxies. Urry et al. (2000)
suggest that, if the mass of the central black hole (BH) in BL Lac objects
correlates with the bulge luminosity of the host galaxy, as is the case in
nearby, inactive galaxies (e.g., Magorrian et al. 1998; Kormendy \& Gebhardt
2001), the Eddington ratios\footnote{The Eddington ratio refers to the ratio
of the bolometric accretion luminosity to the Eddington luminosity,
$L_{\rm Edd} \approx 1.3\times 10^{38} \mbh$ erg~s$^{-1}$, where $\mbh$ is the 
BH mass in units of $M_{\odot}$.} in  BL Lac objects must span a large range.
A number of authors have suggested that the radio power in radio-loud quasars
is controlled by the BH mass (e.g., Laor 2000; Lacy et al. 2001; McLure \&
Dunlop 2001; but see Ho 2002).  If the host galaxies of BL Lac objects
obey the BH mass/bulge luminosity relation, then the peak frequencies are
evidently not controlled by the BH mass (Urry et al. 2000).  This conclusion
has been strengthened by the work of Falomo, Kotilainen \& Treves (2002), Wu,
Liu, \& Zhang (2002), and Barth, Ho, \& Sargent (2002a, b), who have obtained
more robust BH mass estimates using the relation between BH mass and the 
stellar velocity dispersion of the bulge (Gebhardt et al. 2000; Ferrarese \& 
Merritt 2000; Tremaine et al.  2002).

If BH mass is not the main parameter, what about the accretion rate?
The importance of the role of accretion rates in BL Lac objects was originally
recognized by Rees et al. (1982), who suggested that optically thin tori may
power their central engines.  A number of authors have recently revisited this 
issue in the context of BL Lac objects (Ghisellini \& Celotti 2001;
\centerline{\includegraphics[angle=-90,width=8.0cm]{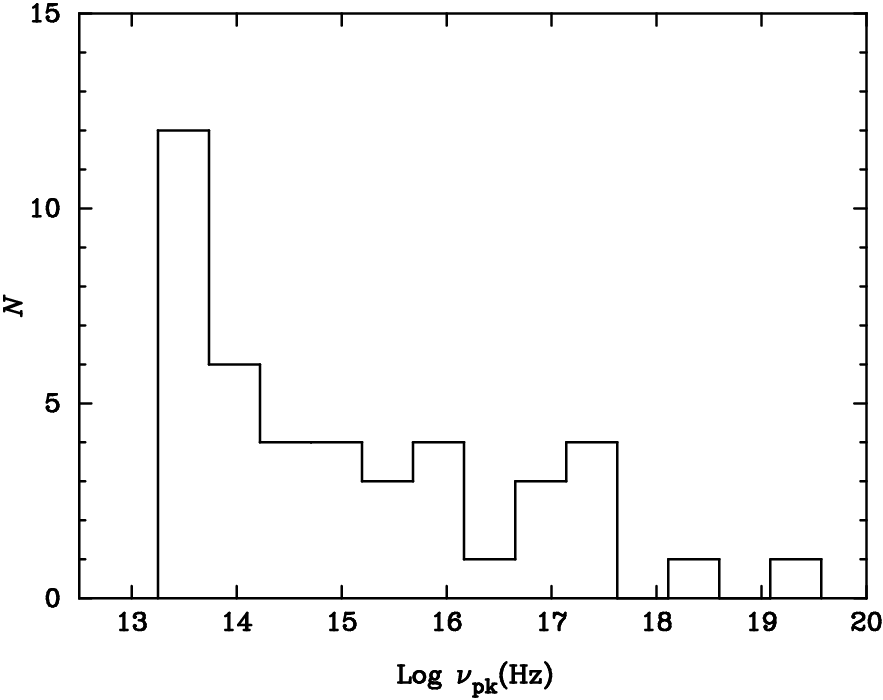}}
\figcaption{\footnotesize
Distribution of peak frequencies for our sample of BL Lac objects. Note that
there is no bimodality in the distribution.
\label{fig1}}
\vskip 0.265cm
\noindent
Maraschi 2001; B\"ottcher \& Dermer 2002; Cavaliere \& D'Elia 2002; Cao 2002)
and more generally radio-loud quasars (Lacy et al. 2001). Using a large sample 
of objects which covers a wide range of nuclear
activity, from nearly inactive systems to classical Seyfert 1 nuclei
and quasars, Ho (2002) shows that the conventional ``radio-loudness'' parameter
is strongly inversely correlated with the mass accretion rate.

This paper presents empirical evidence that the SEDs of BL Lac objects
depend on the accretion rate.

\section{THE SAMPLE}

The sample of BL Lac objects is chosen according to the availability of
information on (1) the luminosity of the host galaxy, (2) the luminosity
of one or more emission lines, and (3) the fitted peak frequency and luminosity
of the SED.  The sample is limited mainly by the host galaxy observations.
Relatively large ground-based optical imaging surveys of BL Lac objects have
been performed by Abraham, McHardy, \& Crawford (1991), Wurtz, Stocke, \& Yee
(1996), and Falomo \& Kotilainen (1999).  More recently, an {\it HST}\ survey
of 132 BL Lac objects was completed by Scarpa et al. (2000) and Urry et
al. (2000).  Whenever possible, we have given preference to the {\it HST}\ data
because of the high resolution and the uniformity of that sample.  Emission-line
fluxes come primarily from the studies of Stickel, Fried, \& K\"uhr (1993),
Rector et al. (2000), and Rector \& Stocke (2001).   Finally, we listed
published values of the peak frequencies and the corresponding peak
luminosities of the SEDs; these are obtained by fitting a logarithmic parabola
to the multiwaveband, usually non-simultaneous, continuum.

The final sample, listed in Table 1, consists of 43 objects.  For convenience,
distance-dependent quantities assume $H_0 = 50$ km~s$^{-1}$~Mpc$^{-1}$ and
$q_0 = 0$.  We note that the original classifications by Urry et al. (2000),
which used $\log F_{\rm keV}/F_{\rm 5GHz}=-5.5$ as the criterion to separate
the two types, contained some HBLs with low peak frequencies ($\pknu$).
Figure 1 shows the distribution of $\pknu$ for our sample.  There is no
obvious bimodality in the distribution; some objects have ``intermediate''
peak frequencies.  For the subsequent analysis, we somewhat arbitrarily
choose $\log \nu_{\rm pk}=15.1$, roughly in the middle of the distribution,
as the boundary between HBLs and LBLs.  With this criterion, our sample has 
17 HBLs and 26 LBLs.

\begin{center}
\footnotesize
\centerline{\sc Table 1}
\centerline{\sc The Sample of BL Lac Objects}
\vglue -0.1cm
\begin{tabular}{cllccccccll}\\ \hline \hline
Source&$~~-\MR$&$\lnupk$&$\llpk$&$\lll$&Ref\\ \hline
0158$+$001&"23.05    &16.87    &44.89&42.56& 1 \\
0205$+$351&"23.33$^a$&17.51    &44.55&42.43& 1 \\
0607$+$710&"24.34    &15.25    &44.04&42.40& 1 \\
0737$+$744&"24.32    &16.16    &45.21&42.65& 1 \\
0922$+$749&"24.64    &17.21    &44.91&43.20& 1 \\
1207$+$394&"24.40    &16.75    &45.24&42.55& 1 \\
1221$+$245&"22.49    &15.27    &44.36&42.78& 1 \\
1229$+$643&"24.07    &17.23    &44.60&43.39& 1 \\
1312$-$422&"23.38$^b$&15.90$^d$&...  &42.45& 1 \\
1407$+$595&"24.78    &15.90    &44.50&42.77& 1 \\
1443$+$635&"23.86$^a$&19.57    &44.41&42.20& 1 \\
1458$+$224&"23.69    &15.40    &45.00&44.30& 1 \\
1534$+$014&"24.21    &16.88    &44.70&42.13& 1 \\
1652$+$398&"24.20$^c$&16.00$^e$&...  &42.20& 2 \\
1757$+$703&"23.63    &17.18    &45.09&42.36& 1 \\
2005$-$489&"23.86    &18.26    &45.47&42.03& 2 \\
2143$+$070&"23.68    &16.34    &44.67&42.50& 1 \\
\hline
0122$+$090&"23.75    &13.47    &44.51&42.56& 1 \\
0235$+$164&<27.25    &13.39    &46.72&44.27& 2 \\
0257$+$342&"24.05    &14.88    &44.24&42.27& 1 \\
0317$+$183&"23.71    &14.36    &44.27&42.60& 1 \\
0419$+$194&"24.01    &13.83    &45.00&42.84& 1 \\
0537$-$441&<26.96    &14.07    &46.62&45.31& 2 \\
0851$+$202&<23.69    &13.72    &46.03&43.92& 2 \\
0954$+$658&<23.23    &14.09    &45.39&44.18& 3 \\
1144$-$379&<24.99    &13.75    &46.54&44.65& 2 \\
1235$+$632&"23.95$^a$&14.88    &44.35&42.65& 1 \\
1308$+$326&<27.00$^c$&13.83    &46.78&44.96& 2 \\
1402$+$041&<23.08    &14.85    &44.64&43.01& 1 \\
1418$+$546&"24.09    &13.85    &45.11&43.59& 2 \\
1514$-$241&"23.54    &15.09$^f$&...  &42.10& 2 \\
1538$+$149&"24.64    &13.56    &45.74&43.50& 2 \\
1552$+$202&"24.39$^a$&13.70    &44.74&43.51& 1 \\
1749$+$096&"23.60    &13.27    &45.78&43.76& 2 \\
1749$+$701&<27.11    &14.43    &46.13&45.22& 2 \\
1803$+$784&<24.55    &13.43    &46.12&44.90& 3 \\
1807$+$698&"23.95    &14.26    &44.06&43.81& 4 \\
1823$+$568&"25.07    &13.65    &46.01&43.40& 4 \\
2007$+$777&"23.89    &13.66    &45.48&43.76& 2 \\
2131$-$021&<27.02    &13.33    &45.67&44.17& 3 \\
2200$+$420&"23.61    &14.25    &44.66&42.80& 2 \\
2240$-$260&<23.83    &13.32    &45.93&44.13& 2 \\
2254$+$074&"24.41    &13.25    &44.99&43.17& 2 \\
\hline
\end{tabular}
\end{center}
\vskip -3.8pt
\parbox{3.3in}
{\footnotesize\baselineskip 9pt
\indent
{\sc Notes:}
(a) Falomo \& Kotilainen 1999;
(b) Falomo \& Ulrich 2000;
(c) Pursimo et al. 2002;
(d) Wolter et al. 1998;
(e) Comastri, Molendi \& Ghisellini 1995;
(f) Landau et al. 1986.
References for emission-line fluxes:
(1) Rector et al. 2000;
(2) Stickel, Fried, \& K\"uhr 1993.
(3) Rector \& Stocke 2001;
(4) Lawrence et al. 1996.
The data for absolute magnitudes are taken from Urry et al. 2000, and the
peak frequencies and luminosities are taken from Sambruna et al.  1996. The
objects above and below the dividing are HBLs and LBLs, respectively.}
\vglue 0.2cm
\normalsize

\section{Statistical Properties}

It is nontrivial to measure the thermal, accretion luminosity from BL Lac
objects because the bulk of their emission comes from nonthermal processes
associated with the jet.  In this study, we adopt the strategy of using the 
observed luminosity of the emission lines to estimate the accretion luminosity 
of the disk.  Emission lines themselves, however, are difficult to measure in 
BL Lac objects because they are both intrinsically weak and heavily diluted by 
the strong featureless continuum.  Nevertheless, they have been detected in a 
number of sources, and our sample was chosen with this application in mind.

In our sample, the most commonly detected emission lines are \mg\
$\lambda$2800, \oo\ $\lambda$3727, H$\beta$, \ooo\ $\lambda$5007, and
H$\alpha$.  To estimate the {\it total}\ luminosity due to emission lines, we
follow the method of Celotti, Padovani \& Ghisellini (1997), who 
compare the luminosity of the observed lines with the fractional 
contribution these lines make to the total line luminosity, as 

\centerline{\includegraphics[angle=-90,width=8.0cm]{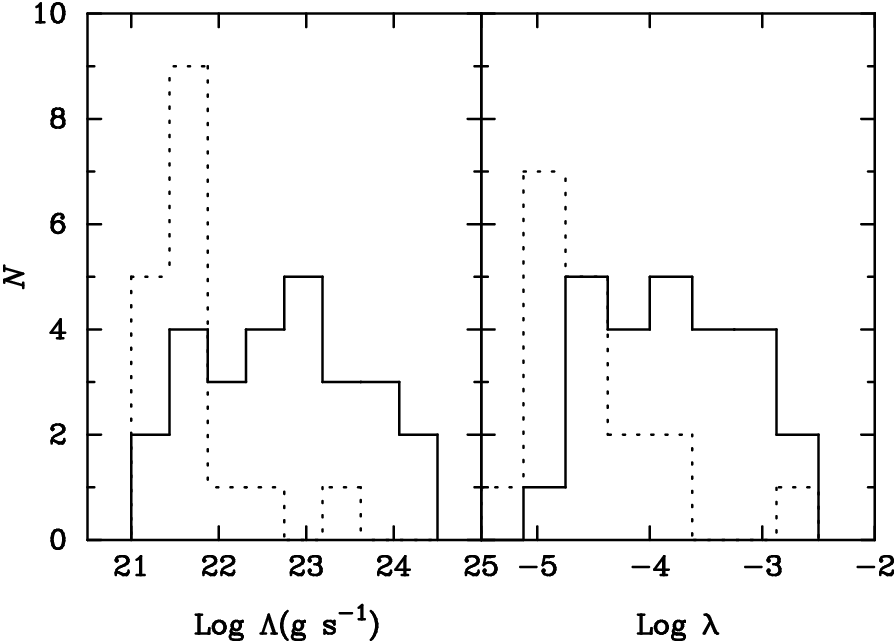}}
\figcaption{\footnotesize
The distribution of accretion rates in BL Lac objects. The dotted histogram
represent the HBLs and solid histograms show the LBLs.
\label{fig2}}
\vskip 0.3cm

\noindent
determined from the relative line
ratios computed by Francis et al. (1991) for a composite quasar spectrum.  As
in Celotti et al. (1997), we assume that the total line luminosity
$\langle L_{\rm lines}\rangle = 556 \langle L_{\rm Ly\alpha}\rangle$.
The line luminosities are given in Table 1. 

With the assumption that the bulk of the line luminosity is photoionized
by the accretion disk (Netzer 1990), $L_{\rm lines}$  should be proportional 
to the total disk luminosity.  We define a ``line accretion rate'' and its 
dimensionless form as follows:
\begin{equation}
\Lambda=\frac{\lline}{c^2};~~~
\lambda=\frac{\lline}{L_{\rm Edd}}.
\end{equation}
The relation between $\lambda$ and the dimensionless accretion rate $\dot{m}$
is given by equation (2).  We determine BH masses from the absolute $R$-band
magnitudes of the host galaxies, using the relation determined by McLure \&
Dunlop (2001) from a study of active galaxies: $\log \mbh =
-0.5\MR-2.96$.  This relation is quite similar, but not identical, to that
given by Kormendy \& Gebhardt (2001) for inactive galaxies.

The above estimates of the accretion rates may carry significant
uncertainties that are difficult to quantify.  While the formal errors 
on the luminosities of individual emission lines are usually quite modest, 
what is less certain is the conversion to total line luminosity, as this 
step implicitly assumes that the relative line strengths in BL Lac objects 
are the same as in typical quasars.  Thus, our values of $L_{\rm lines}$, 
and hence $\Lambda$ and $\lambda$, are potentially affected by systematic 
errors.   The values of $\lambda$ are additionally affected by the relatively 
large scatter in the BH mass versus bulge luminosity relation ($\sim$0.59 dex; 
McLure \& Dunlop 2001).  Despite the potentially large uncertainties on 
any individual measurement of $\Lambda$ or $\lambda$, and their inferred 
accretion rates, we would like to emphasize that the {\it relative}\ values of 
these quantities, such as those deduced for HBLs and LBLs discussed below, 
should be more robust.

\subsection{Accretion Rates}

Figure 2 shows that the accretion rates vary over a rather wide range in LBLs,
whereas they are more restricted in HBLs.  The mean values of the accretion
rates are $\langle\log \Lambda\rangle=22.70\pm 0.91$(rms dispersion) and
$\langle\log \lambda\rangle=-3.82\pm0.68$ in LBLs and
$\langle\log \Lambda\rangle=21.69\pm 0.56$ and $\langle\log \lambda\rangle =
-4.48\pm0.62$ in HBLs. These results directly show that the accretion rate may
control the SEDs of BL Lac objects.  It is important to note that the
distributions of $\Lambda$ and $\lambda$ significantly overlap between
HBLs and LBLs.  This clearly reflects the continuous distribution of peak 

\centerline{\includegraphics[angle=-90,width=8.cm]{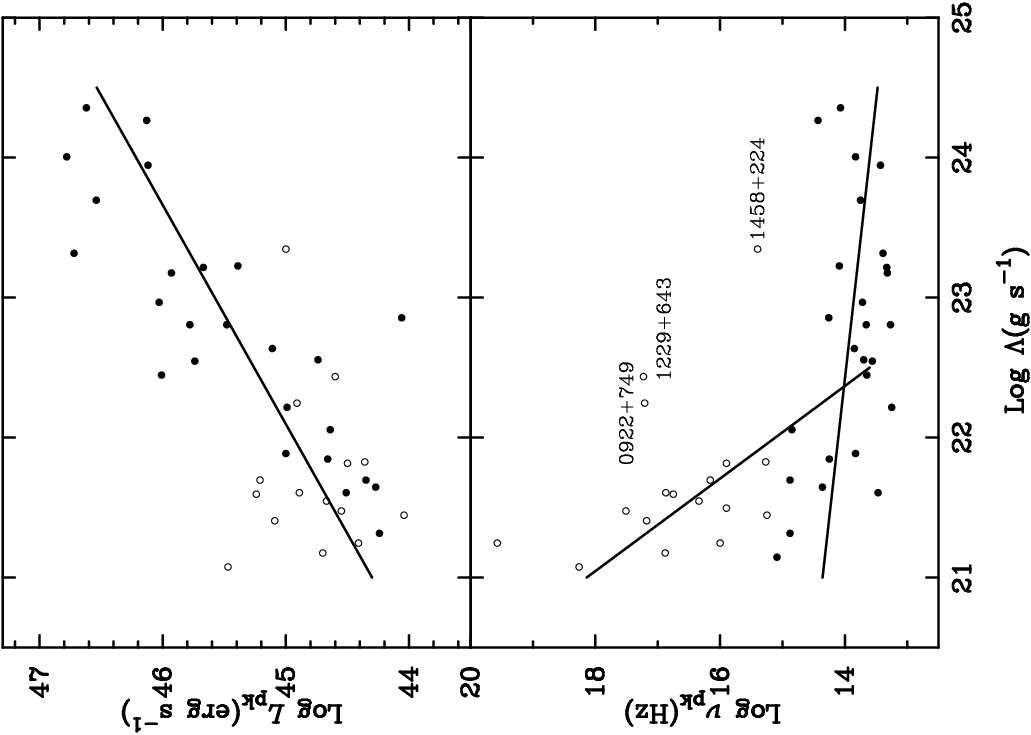}}
\figcaption{\footnotesize
The correlations among the parameters $\pknu$, $\pkl$, and $\Lambda$.
The solid and open symbols denote LBLs and HBLs, respectively.
\label{fig3}}
\vglue 5mm

\noindent
frequencies in Figure 1.

With the assumption that the line luminosity traces the disk luminosity, 
the distribution of $\lambda$ implies that the vast majority of BL Lac objects
are highly sub-Eddington systems.  In such a regime, accretion is thought to
proceed via an optically thin ``advection-dominated'' accretion flow
(ADAF; see Narayan, Mahadevan, \& Quataert 1998).  From Mahadevan 
(1997), the total luminosity from an ADAF is 
$L_{\rm disk}\propto \alpha^{-2}\mbh\dot{m}^2$, where
$\dot{m}=\dot{M}/\dot{M}_{\rm Edd}$,
$\dot{M}_{\rm Edd}=1.39\times 10^{18}\mbh/\eta_{-1}\,{\rm (g~s^{-1})}$,
$\eta=0.1\eta_{-1}$ is the accretion efficiency, and $\alpha$ is the
canonical viscosity parameter.  For a reprocessing efficiency $\xi$,
$\lline=\xi L_{\rm disk}$ and
\begin{equation}
\dot{m}=2.17\times 10^{-2}\alpha_{0.3}\xi_{-1}^{-1/2}\lambda_{-4}^{1/2},
\end{equation}
where $\alpha_{0.3}=\alpha/0.3$, $\xi_{-1}=\xi/0.1$ and $\lambda_{-4}=
\lambda/10^{-4}$.  We assume that $\xi$ is of the same order as the covering
factor of the line-emitting clouds ($\sim 0.1$) since the total energy emitted
by the clouds is simply the energy absorbed by them (Netzer 1990).

An optically thin ADAF can only exist below a critical value of the accretion
rate $\dot{m}_{\rm c} \approx \alpha^2$ (Narayan et al. 1998).  For a
plausible value of $\alpha=0.3$, $\dot{m}_{\rm c} \approx 0.1$.  We find
$\langle \log \dot{m}\rangle=-1.5\pm 0.34$
for LBLs and $\langle \log \dot{m}\rangle=-2.0\pm 0.31$ for HBLs.
This indicates that BL Lac objects are in the ADAF regime, and that LBLs
typically have significantly higher accretion rates than HBLs.  These results
are in qualitative agreement with the findings of Ghisellini \& Celotti (2001)
and Cavaliere \& D'Elia (2002).
We note that equation (2) may not be applicable to some of the LBLs with the
highest values of $\lambda$ because they have $\dot{m}\sim \dot{m}_{\rm c}$;
these objects straddle the (uncertain) transition from a standard disk to
an ADAF (cf. Cao 2002).

\subsection{Accretion Rates and SEDs}

Figure 3 shows the dependence of the peak frequency ($\pknu$) and peak
luminosity ($\pkl$) on $\Lambda$.  We find that $\pkl$ correlates strongly
with $\Lambda$; the Pearson's correlation coefficient is $r=0.76$,
significant at a level of $>$99.999\%.  Treating $\Lambda$ as the independent
variable, a linear regression fit gives
\begin{equation}
\log \pkl ({\rm erg~s^{-1})}=(30.88\pm 1.99)+(0.64\pm 0.09)\log\Lambda.
\end{equation}
An obvious interpretation of the $\pkl-\Lambda$ correlation is that the
jet power is directly coupled to the mass accretion rate of the disk.
The disk-jet ``connection'' in active nuclei has been discussed in a number
of contexts (e.g., Rawlings \& Saunders 1991; Falcke \& Biermann 1995;
Xu, Livio \& Baum 1999; Ho \& Peng 2001), and the present correlation 
provides additional support for this picture because the peak luminosity of 
the SED of a BL Lac object represents most of the radiated energy from the jet.

The relation between $\pknu$ and $\Lambda$ is more complex.  The HBL group
and the LBL group each seems to define its own sequence.  After excluding
three strongly deviant points (0922+749, 1229+643 and 1458+224)\footnote{The 
emission-line equivalent widths of 0922+749 and 1229+643 formally exceed the 
(admittedly arbitrary) limit of 5 \AA. The value of $\pknu$ for 0922+749 was 
obtained from fitting only three points (Sambruna et al. 1996) and is 
therefore quite uncertain. There is no obvious explanation for 1458+224.}, 
HBLs appear to show a moderately strong correlation with $r=-0.6$ at a 
significance of 98\%. The best-fit linear regression gives
\begin{equation}
\log \pknu(\rm Hz)=(81.65\pm 25.61) -(3.02\pm 1.19)\log \Lambda.
\end{equation}
The peak frequency formally scales with the accretion rate as
$\pknu\propto \Lambda^{-3}\propto \dot{M}^{-6}$, but we note that this is
somewhat uncertain because of the limited dynamic range in $\Lambda$.
For LBLs, we have
\begin{equation}
\log \pknu (\rm Hz)=(19.66\pm 2.55)-(0.25\pm 0.11)\log \Lambda,
\end{equation}
with $r=-0.42$ and a significance of $97\%$.  This implies $\pknu\propto
\Lambda^{-0.25} \propto \dot{M}^{-0.5}$, much shallower than in HBLs.
Bearing in mind the above caveat on the HBLs, the different slopes of the
$\pknu-\Lambda$ relation may indicate differences in the physical processes
associated with the jet, as discussed in the next section.  It is interesting
to note that HBLs and LBLs also delineate different loci, separated by a sharp
break, when $\pknu$ is plotted against the radio luminosity (Costamante et al.
2001).

\section{Discussion}

According to Xu et al. (1999), the accretion rate through the disk ($\dot{M}$)
is related to the mass flux into the jet ($\dot{M}_{\rm jet}$) by
\begin{equation}
\dot{M}_{\rm jet}/\dot{M}\sim \left(H/R\right)^{2\delta-3},
\end{equation}
where $H$ is the height of the disk at radius $R$ and $\delta$ is a constant
from 1.7$-$3.4. From the distribution of $\lambda$, we inferred that both HBLs
and LBLs contain ADAFs.  Since the structure of ADAFs is quasi-spherical,
$H/R\sim 1$ or $\dot{M}_{\rm jet}\sim \dot{M}$. As Blandford \& Begelman
(1999) argued, ADAFs inevitably lead to outflows.  If energy is extracted from
the spin of the BH via the Blandford-Znajek (1977) mechanism, $L_{\rm BZ}=
\epsilon_{\rm BZ}\dot{M}c^2$, where $\epsilon_{\rm BZ}=1.8\times 10^{-2}$ and
the BH spin is assumed to be maximal (Armitage \& Natarajan 1999). The BZ
process has been strengthened by a recent observation of Fe K$\alpha$ line
profile in MCG 6-30-15 (Wilms et al. 2001). This energy
will be channeled into two parts: the kinetic luminosity
$L_{\rm kin}=\dot{M}_{\rm jet}\Gamma c^2/2$ ($\Gamma$ is the Lorentz factor
of the jet) and the radiative luminosity $L_{\rm rad}^{\prime}$ (seen in the
comoving frame of the jet; e.g., Sikora et al. 1997).  Here, we explore the
possibility that the relative partition of the spin energy determines the 
shape of the SED.  In particular, we investigate whether the different 
functional dependence of $\nu_{\rm pk}$ on $\Lambda$ for HBLs and LBLs can be 
understood in terms of different radiative efficiencies in the jet.

Let us assume that the energy density of the random component of the magnetic
field is proportional to the energy density of the relativistic electrons in
the jet. Then, $B^2\propto \langle E_e\rangle n_e$, where $\langle E_e\rangle$
is the mean energy and $n_e$ is the number density of the relativistic
electrons.  With the help of $\dot{M}_{\rm jet}\sim R_0^2 n_e m_p c$, where
$R_0$ is the radius of the jet and $m_p$ is the proton mass, we have
$B\propto \dot{M}_{\rm jet}^{1/2}R_0^{-1}$ and 
$n_e\propto \dot{M}_{\rm jet}R_0^{-2}$.  If the relativistic electrons
have a power-law distribution of energies $N=N_0E_e^{-\gamma}$, the emissivity
at the peak frequency $\nu_{\rm pk}^{\prime}$ in the jet comoving frame 
is given by $\epsilon_{\rm pk}^{\prime}\propto 
N_0B^{(\gamma+1)/2}\nu_{\rm pk}^{{\prime}^{(1-\gamma)/2}}\propto 
n_eB^{(\gamma+1)/2}\nu_{\rm pk}^{{\prime}^{(1-\gamma)/2}}$ 
(Pacholczyk 1970), and it follows that 
$\epsilon_{\rm pk}^{\prime}\propto R_0^{-(5+\gamma)/2}
                                   \dot{M}_{\rm jet}^{(\gamma+5)/4} 
                                   \nu_{\rm pk}^{{\prime}^{(1-\gamma)/2}}$.
The radiative luminosity in the jet comoving frame will then be given by
$L_{\rm rad}^{\prime}\propto
R_0^3\left(\nu_{\rm pk}^{\prime}\epsilon_{\rm pk}^{\prime}\right)
\propto R_0^{(1-\gamma)/2}\dot{M}_{\rm jet}^{(5+\gamma)/4}
\nu_{\rm pk}^{{\prime}^{(3-\gamma)/2}}$.  The ratio of the 
radiative to the kinetic luminosity will be proportional to 
$\Gamma^{-1}\dot{M}_{\rm jet}^{(1+\gamma)/4} 
\nu_{\rm pk}^{{\prime}^{(3-\gamma)/2}}$, which suggests that the higher peak 
frequencies lead to the higher radiative efficiencies.

For HBLs, we consider the limiting case that most of the energy is radiated.
By setting $L_{\rm rad}^{\prime}\sim L_{\rm BZ}$, we have
$\nu_{\rm pk}^{\prime}\propto \dot{M}^{-(\gamma+1)/2(3-\gamma)}$, and the
observed peak frequency is
\begin{equation}
\nu_{\rm pk}={\cal D}\nu_{\rm pk}^{\prime}\propto
\dot{M}^{-(\gamma+1)/2(3-\gamma)} \propto \Lambda^{-(\gamma+1)/4(3-\gamma)}.
\end{equation}
Here, we have assumed that the Doppler factor ${\cal D}$ does not vary
strongly with $\dot{M}$.
We see that $\pknu$ depends sensitively on $\Lambda$
for a given $\gamma$. For a typical value of $\gamma \approx 2.6-2.7$
(Kirk, Melrose \& Priest 1994), we get $\pknu\propto \Lambda^{-2.3}$ to
$\Lambda^{-3.1}$. This is consistent with the empirical results for HBLs
(eq. 4).  Despite the apparent agreement, however, we must regard this
result with some caution because of the extreme requirement that the bulk
of the jet power emerge as radiative energy.  Celotti \& Fabian (1993) find 
that most jet sources do not radiate efficiently, although there are some
objects with high radiative efficiencies (see their Fig. 2).

Since LBLs have lower peak frequencies, their radiative efficiencies are lower,
and we postulate that in this case the kinetic luminosity dominates over the
radiative luminosity. The maximum energy of the relativistic electrons will be
given by the balance between energy loss and gain. The acceleration timescale
is $\tau_{\rm acc}\sim l(E_e)/c\propto B^{-1}E_e^n$, where $l(E_e)=l_0E_e^n$ is
the mean free path and $n=1$ in the Bohm limit (Inoue \& Takahara 1996). The
timescale for energy loss due to inverse Compton scattering of external
photons is $\tau_{\rm loss}\propto (u_{\rm ext}E_e)^{-1}$, where the energy
density of external photons $u_{\rm ext}\propto L_{\rm disk}$ (Sikora,
Begelman \& Rees 1994). We thus have
$E_{e,\rm max}\propto \left(B/L_{\rm disk}\right)^{1/(n+1)}$, and the peak
frequency is
\begin{equation}
\nu_{\rm pk}\propto {\cal D}BE_{e,\rm max}^2\propto \dot{M}^{(n-5)/2(n+1)}
            \propto\Lambda^{(n-5)/4(n+1)},
\end{equation}
where we used $L_{\rm disk}\propto \dot{M}^2$. For the Bohm limit ($n=1$)
or not very efficient acceleration (e.g., $n=2$), the relation will be
$\pknu\propto \Lambda^{-0.25}$ to $\Lambda^{-0.50}$.  This, too, is roughly
consistent with the empirical correlation for LBLs (eq. 5).  In this scenario,
which is similar to that discussed by Ghisellini et al. (1998) and B\"ottcher 
\& Dermer (2002), the peak frequencies in LBLs are determined by cooling of 
relativistic electrons due to inverse Compton scattering of photons from the 
disk or reflected off of the broad-line emitting clouds.

Finally, we would like to note that our analysis hinges on the assumption that
the emission lines in BL Lac objects are photoionized by the disk, such that
line luminosity reliably traces the accretion luminosity.  Based on
spectrophotometric monitoring of a sample of high-redshift, high-luminosity
active galactic nuclei, P\'erez, Penston \& Moles (1989) find that the
emission lines vary on timescales shorter than expected from photoionization
models.  They suggest that the broad-line region in these objects may be
anisotropic.  
However, the variability behavior of the broad optical
line H$\alpha$ in BL Lac itself appears to support that ``H$\alpha$ emission
could be powered by thermal radiation from an accretion disk without
significantly affecting the shape ot polarization of the optical continuum'' 
(Corbett et al. 1996, 2000).  If it turns out that a significant
fraction of the line luminosity is powered by the jet instead of the disk,
then clearly the interpretation of Figure 3 needs to be reevaluated.
However, the conclusion that BL Lac objects are highly sub-Eddington,
advection-dominated systems would not change; in fact, it would only be
strengthened because the inferred accretion luminosity would be even
lower than we have assumed.

\section{Conclusions}

Using a sample of BL Lac objects with information on their host
galaxies and emission-line luminosities, we show that LBLs have significantly
higher accretion rates and Eddington ratios than HBLs.  Both classes are
highly sub-Eddington systems that may be accreting via an ADAF.  We find that 
the peak luminosity of the SED correlates significantly with the 
accretion rate, lending 
strong support to the idea that the jet and accretion disk are 
closely coupled.  

Furthermore, we present evidence that the peak frequency of the SED also 
correlates with the accretion rate, although the functional dependence between 
the two parameters appears to be quite different between HBLs and LBLs.  
We argue that these empirical trends can be qualitatively explained by 
invoking different radiative efficiencies in the jets of the two classes of 
objects.  HBLs, which have lower accretion rates, evidently manage to convert 
more of their jet kinetic power into radiation.  By contrast, in LBLs, which 
have higher accretion rates, a greater fraction of the jet power remains in 
kinetic form.  However, the underlying causal connection between the radiative 
efficiency of the jet and the accretion rate, and the manner in which energy 
is channeled into relativistic electrons (Ghisellini, Celotti \& Costamante
2002), remain unclear.

It would be of interest to perform spectrophotometric monitoring of 
nearby BL Lac objects in order to establish more conclusively the origin of 
their broad emission lines.

\acknowledgements
{J.M.W. acknowledges support from the Alexander von Humboldt Foundation,
the ``Hundred Talents Program of CAS'', the Special Funds for Major State
Basic Research Project and NSFC.}


\begin{thebibliography}{99}
\bibitem{}
Abraham, R.~G., McHardy, I.~M., \& Crawford, C.~S. 1991, MNRAS, 252, 482

\bibitem{}
Armitage, P.~J., \& Natarajan, P. 1999, ApJ, 523, L7

\bibitem{}
Barth, A.~J., Ho, L.~C., \& Sargent, W.~L.~W. 2002a, ApJ, 566, L13

\bibitem{}
------. 2002b, ApJ, submitted

\bibitem{}
Blandford, R.~D., \& Begelman, M.~C. 1999, MNRAS, 303, L1

\bibitem{}
Blandford, R.~D., \& Znajek, R.~L. 1977, \mnras, 179, 433


\bibitem{}
B\"ottcher, M., \& Dermer, C.~D. 2002, \apj, 564, 86

\bibitem{}
Caccianiga, A., Maccacaro, T., Wolter, A., Della Ceca, R., \& Gioia, I.~M.
1999, \apj, 513, 51

\bibitem{}
Cao, X. 2002, \apj, 570, L13

\bibitem{}
Cavaliere, A. \& D'Elia, V. 2002, ApJ, 571, 226

\bibitem{}
Celotti, A., \& Fabian, A.~C. 1993, MNRAS, 264, 228

\bibitem{}
Celotti, A., Padovani, P. \& Ghisellini G. 1997, MNRAS, 286, 415

\bibitem{}
Comastri, A., Molendi, S. \& Ghisellini, G. 1995, MNRAS, 277, 297

\bibitem{}
Corbett, E.~A., Robinson, A., Axon, D.~J., \& Hough, J.~H. 2000, MNRAS, 
311, 485

\bibitem{}
Corbett, E.~A., Robinson, A., Axon, D.~J., Hough, J.~H., Jeffries, R.~D., 
Thurston, M.R. \& Young, S. 1996, MNRAS, 281, 737 

\bibitem{}
Costamante, L., et al. 2001, A\&A, 371, 512

\bibitem{}
Falcke, H., \& Biermann, P.~L. 1995, A\&A, 293, 665

\bibitem{}
Falomo, R., \& Kotilainen, J.~K., 1999, A\&A, 352, 85

\bibitem{}
Falomo, R., Kotilainen, J.~K., \& Treves, A. 2002, \apj, 569, L35

\bibitem{}
Falomo, R., \& Ulrich, M.-H. 2000, A\&A, 357, 91

\bibitem{}
Ferrarese, L., \& Merritt, D. 2000, ApJ, 539, L9

\bibitem{}
Francis, P.~J., Hewett, P.~C., Foltz, C.~B., Chaffee, F.~H., Weymann, R.~J.,
\& Morris, S.~L. 1991, \apj, 373, 465


\bibitem{}
Gebhardt, K., et al. 2000, \apj, 539, L13

\bibitem{}
Ghisellini, G., \& Celotti A. 2001, A\&A, 379, L1

\bibitem{}
Ghisellini, G., Celotti A., Costamante, L.,  2002, A\&A, 386, 833

\bibitem{}
Ghisellini, G., Celotti, A., Fossati, G., Maraschi, L., \& Comastri, A.
1998, MNRAS, 301, 451

\bibitem{}
Giommi, P., \& Padovani, P. 1994, MNRAS, 268, L51

\bibitem{}
Ho, L.~C. 2002, ApJ, 564, 120

\bibitem{}
Ho, L.~C., \& Peng, C.~Y. 2001, \apj, 555, 650

\bibitem{}
Inoue, S., \& Takahara, F. 1996, ApJ, 463, 555

\bibitem{}
Jannuzi, B.~T., Smith, P.~S. \& Elston, R. 1994, ApJ, 428, 130

\bibitem{}
Kirk, J., Melrose, D.~B., \& Priest, E.~R. 1994, in Plasma Astrophysics, ed. 
A.~O. Benz, \& T.~J.-L. Courvoisier (Heidelberg: Springer-Verlag)

\bibitem{}
Kormendy, J., \& Gebhardt, K. 2001, in The 20th Texas Symposium on Relativistic
Astrophysics, ed. H. Martel \& J.~C. Wheeler (New York: AIP), 363

\bibitem{}
Lacy, M., Laurent-Meuleisen, S.~A., Ridgway, S.~E., Becker, R.~H, \&
White, R.~L. 2001, \apj, 551, L17:

\bibitem{}
Landau, R., et al. 1986, ApJ, 308, 78

\bibitem{}
Laor, A. 2000, ApJ, 543, L111

\bibitem{}
Laurent-Muehleisen, S.~A., Kollgaard, R.~I., Ciardullo, R., Feigelson,
E.~D., Brinkmann, W., \& Siebert, J. 1998, \apjs, 118, 127

\bibitem{}
Laurent-Muehleisen, S.~A., Kollgaard, R.~I., Moellenbrock, G.~A., \&
Feigelson, E.~D. 1993, AJ, 106, 875

\bibitem{}
Lawrence, C.~R., Zucker, J.~R., Readhead, A.~C.~S., Unwin, S.~C., Pearson, 
T.~J., \& Xu, W. 1996, ApJS, 107, 541

\bibitem{}
Magorrian, J., et al. 1998, AJ, 115, 2285

\bibitem{}
Mahadevan, R. 1997, ApJ, 477, 585

\bibitem{}
Maraschi, L. 2001, in The 20th Texas Symposium on Relativistic
Astrophysics, ed. H. Martel \& J.~C. Wheeler (New York: AIP), 409

\bibitem{}
McLure, R.~J., \& Dunlop, J.~S. 2001, \mnras, 327, 199

\bibitem{}
Narayan, R., Mahadevan, R., \& Quataert, E. 1998, in The Theory of Black Hole
Accretion Discs, ed.  M. A. Abramowicz, G. Bj\"{o}rnsson, \& J. E. Pringle
(Cambridge: Cambridge Univ. Press), 148

\bibitem{}
Netzer, H. 1990, in Active Galactic Nuclei, ed. R.~D. Blandford, H. Netzer, \&
L. Woltjer (Berlin: Springer), 57

\bibitem{}
Pacholczyk, A.~G. 1970, Radio Astrophysics (San Francisco: W.~H. Freeman and 
Company)

\bibitem{}
P\'erez, E., Penston, M.V., \& Moles, M., 1989, MNRAS, 239, 75

\bibitem{}
Perlman, E.~S., Padovani, P., Giommi, P., Sambruna, R., Laurence, R.~J.,
Tzioumis, A., \& Reynolds, J. 1998, \aj, 115, 1253

\bibitem{}
Perlman, E.~S. \& Stocke, J.~T., 1993, ApJ, 406, 430

\bibitem{}
Pursimo, T., Nilsson, K., Takalo, L.~O., Sillanp\"a\"a, A., Heidt, J., \&
Pietil\"a, H. 2002, A\&A, 381, 810

\bibitem{}
Rawlings, S., \& Saunders, R. 1991, \nat, 349, 138

\bibitem{}
Rector, T.~A. \& Stocke, J.~T. 2001, AJ, 122, 565

\bibitem{}
Rector, T.~A., Stocke, J.~T., Perlman, E.~S., Morris, S.~L., \& Gioia,
I.~M. 2000, \aj, 120, 1626

\bibitem{}
Rees, M.~J., Begelman, M.~C., Blandford, R.~D., \& Phinney, E.~S. 1982, Nature,
295, 17

\bibitem{}
Sambruna, R.~M., Maraschi, L., \& Urry, C.~M. 1996, ApJ, 463, 444

\bibitem{}
Scarpa, R., Falomo, R., \& Pian, E., \& Treves, A. 1995, A\&A, 303, 730

\bibitem{}
Scarpa, R., Urry, C.~M., Falomo, R., Pesce, J.~E., \& Treves, A. 2000,
\apj, 532, 740

\bibitem{}
Sikora, M., Begelman, M.~C. \& Rees, M.~J. 1994, ApJ, 421, 153

\bibitem{}
Sikora, M., Madejski, G., Moderski, R. \&  Poutanen, J. 1997, \apj, 484, 108


\bibitem{}
Smith, P.S. \& Sitko, M.L., 1991, ApJ, 383, 580 

\bibitem{}
Stickel, M., Fried, J.~W., \& K\"uhr, H. 1993, A\&AS, 98, 393

\bibitem{}
Tremaine, S., et al. 2002, \apj, in press (astro-ph/0203468)


\bibitem{}
Urry, C.~M., \& Padovani, P. 1995, PASP, 107, 803

\bibitem{}
Urry, C.~M., Scarpa, R., O'Dowd, M., Falomo, R., Pesce, J.~E., \& Treves,
A. 2000, \apj, 532, 816


\bibitem{}
Wilms, J., et al., 2001, MNRAS, 328, L7

\bibitem{}
Wolter, A., et al. 1998, A\&A, 335, 899

\bibitem{}
Wu, X.-B., Liu, F.~K., \& Zhang, T.~Z. 2002, A\&A, 389, 742

\bibitem{}
Wurtz, R., Stocke, J.~T., \& Yee, H.~K.~C. 1996, ApJS, 103, 109

\bibitem{}
Xu, C., Livio, M., \& Baum, S.~A. 1999, AJ, 118, 1169

\end{thebibliography}
\end{document}